\documentclass[12pt]{article}
\usepackage{times}
\usepackage{geometry}
\usepackage[utf8]{inputenc}
\usepackage{enumitem,amssymb}
\usepackage[numbers,sort]{natbib}
\usepackage{graphics,graphicx}
\usepackage{color}
\usepackage[table]{xcolor}
\usepackage{colortbl} 
\usepackage{tabularx}
\usepackage{caption}
\usepackage{paralist} 
\usepackage[colorlinks=true, citecolor=blue, linkcolor=black, urlcolor=black]{hyperref}
\usepackage{ragged2e}
\newlist{thematicfjhkjf}{itemize}{8}
\usepackage{pifont}
\usepackage[para]{footmisc}

\newcommand{\spbefore}{-0.49cm}
\newcommand{\spafter}{-0.25cm}

\usepackage{wrapfig}
\usepackage{graphics,graphicx}
\usepackage{color}
\usepackage{hyperref}
\usepackage{caption}

\def\apj{\rm ApJ}
\def\apjl{\rm ApJL}
\def\apjs{\rm ApJS}

\def\mnras{\rm MNRAS}

\def\nat{\rm Nature}

\def\aap{\rm A\&A}

\begin{document}
\raggedright
\huge
Astro2020 Science White Paper \linebreak

High-Energy Polarimetry - a new window to probe extreme physics in AGN jets \linebreak
\normalsize

\noindent \textbf{Thematic Areas:} \hspace*{60pt} $\square$ Planetary Systems \hspace*{10pt} $\square$ Star and Planet Formation \hspace*{20pt}\linebreak
$\square$ Formation and Evolution of Compact Objects \hspace*{31pt} \makebox[0pt][l]{$\square$}\raisebox{.0ex}{\hspace{-0.05em}$\checkmark$} Cosmology and Fundamental Physics \linebreak
  $\square$  Stars and Stellar Evolution \hspace*{1pt} $\square$ Resolved Stellar Populations and their Environments \hspace*{40pt} \linebreak
  $\square$    Galaxy Evolution   \hspace*{45pt}           \makebox[0pt][l]{$\square$}\raisebox{.0ex}{\hspace{-0.05em}$\checkmark$} Multi-Messenger Astronomy and Astrophysics \hspace*{65pt} \linebreak
 \linebreak
\textbf{Principal authors:} B.~Rani (NASA GSFC, USA; Email: bindu.rani@nasa.gov; Phone: +1 301.286.2531), H.~Zhang (Purdue University, USA; phitlip2007@gmail.com), S.~D.~Hunter (NASA GSFC, USA; stanley.d.hunter@nasa.gov), F.~Kislat (University of New Hampshire, USA; fabian.kislat@unh.edu), M.~B\"ottcher (North-West University, South Africa; Markus.Bottcher@nwu.ac.za)
 \linebreak
\textbf{Co-authors:} {\small J.~E.~McEnery (NASA GSFC, USA), D.~J.~Thompson (NASA GSFC, USA), D.~Giannios (Purdue University, USA), F.~Guo (Los Alamos National Laboratory, USA), H.~Li (Los Alamos National Laboratory, USA), M.~Baring (Rice University, USA), I.~Agudo (IAA-CSIC, Spain), S.~Buson (Univ. of Wuerzburg), M.~Petropoulou (Princeton University, USA), V.~Pavlidou (University of Crete, Greece), E.~Angelakis (MPIfR, Germany), I.~Myserlis (MPIfR, Germany), Z.~Wadiasingh (NASA GSFC, USA),  R.~M.~Curado da Silva (LIP, Portugal), P.~Kilian (LANL, USA), S.~Guiriec (George Washington University/NASA GSFC, USA), V.~V.~Bozhilov (Sofia University, Bulgaria), J.~Hodgson (KASI, Republic of Korea) S.~Ant{\'o}n (CIDMA/Dep Fisica - Univ.Aveiro, Portugal), D.~Kazanas (NASA GSFC, USA), P. Coppi (Yale University, USA), T. Venters (NASA GSFC, USA), F.~Longo (University of Trieste and INFN Trieste, Italy), E.~Bottacini (University of Padova), R.~Ojha (UMBC/NASA GSFC, USA), B.~Zhang (University of Nevada, USA), S.~Ciprini (NFN Tor Vergata Rome and SSDC ASI, Italy), A.~Moiseev (UMD/NASA GSFC, USA), C.~Shrader (NASA GSFC)}
  \linebreak
\textbf{Abstract:}
The constantly improving sensitivity of ground-based and space-borne observatories has made possible the detection of high-energy emission 
(X-rays and $gamma$-rays) from several thousands of extragalactic sources. Enormous progress has been made in measuring the continuum flux enabling us to perform imaging, spectral and timing studies. An important remaining challenge for 
high-energy astronomy is measuring polarization. The capability to measure polarization is being realized currently at X-ray energies (e.g.\ with IXPE), 
and sensitive gamma-ray telescopes capable of measuring polarization, such as AMEGO, AdEPT, e-ASTROGAM, etc., are being developed. These future gamma-ray telescopes will probe the radiation mechanisms and magnetic fields of relativistic jets from active galactic nuclei at spatial scales much smaller than the angular resolution achieved with continuum observations of the instrument. In this white paper, we  discuss the scientific potentials of high-energy polarimetry, especially {\bf gamma-ray polarimetry}, including the theoretical implications, and observational technology advances being made. In particular, we will explore the primary scientific opportunities and wealth of information expected from synergy of multi-wavelength polarimetry that will be brought to multi-messenger astronomy. 

\pagebreak

\justifying
\vspace{\spbefore}
\section{Background and Introduction}
\vspace{\spafter}
Advances in high-energy (X-ray to $\gamma$-ray) spectral and timing studies in the past few decades have revolutionized our understanding of the Universe. In particular, active galactic nuclei (AGN) with their relativistic jets pointing along our line-of-sight, called blazars, are found to be among the most violent objects in the Universe and also the most numerous objects in the extragalactic $\gamma$-ray sky \citep{3lac2015}. {\it 
A regime yet to be explored is the high-energy polarization.} High-energy polarimetry provides two additional observables, i.e.\ polarization fraction ({\bf PF}) and polarization angle ({\bf PA}), adding invaluable constraints and insights on the extreme physical processes and morphology of AGN jets.

Blazar double-hump shaped spectral energy distributions (SEDs) are characterized by a low-energy component spanning from radio to soft X-ray and a high-energy component from X-rays up to TeV $\gamma$-rays \citep{hayashida2015, rani2013}. 
They exhibit intense variability across the entire SED time scales as short as a few minutes, implying violent particle acceleration \citep{hayashida2015, rani2013, ackermann2016, abdo2011}. Given their luminous $\gamma$-ray emission and fast variability, blazars have long been suspected as the prime  extragalactic accelerators of cosmic rays. The recent very high energy neutrino event, IceCube-170922A, coincident with the flaring $\gamma$-ray blazar, TXS~0506+056, strongly supports that AGN jets are extragalactic sources of neutrinos \citep{icecube2018}. This novel discovery marks the beginning of multi-messenger astronomy of AGN jets, which will be a top priority in the next decade. {\it To leverage the multi-messenger observations and explore the extreme physics in AGN jets}, we need to understand both electromagnetic and neutrino signatures from blazars: {\bf What makes the high-energy radiation and neutrinos from blazars? How do blazar jets accelerate particles?}

Radio and optical polarimetry have been successful in understanding the low-energy blazar spectral component. The observed high PF has unambiguously pinpointed the non-thermal electron synchrotron emission to be the dominant radiation mechanism in the low-energy spectral component \citep{rani2017,  blinov2018, marscher2008, bhatta2015, magic2018}. Very long baseline interferometry (VLBI) has delivered high resolution polarized radio images of AGN jets very close to the central engine, revealing the overall jet structure and evolution 
\citep{rani2018, casadio2019, marscher2008, bondi2004}. These observations have found that blazars can become very active when a powerful radio outburst emerges, often with significant radio polarization variations \citep{orienti2013, raiteri2013, schinzel2012, agudo2011, rani2018, 3c120_paper, jorstad2013, rani2015}, indicating the location of efficient energy dissipation and the magnetic field evolution therein. In the last decade, simultaneous optical polarization monitoring with $\gamma$-rays has became mature. Violent optical polarization variations, in particular the PA swings, are often accompanied by intense multi-wavelength flares \citep{abdo2010, marscher2010, blinov16, blinov2018}, implying that the magnetic field is also involved in particle acceleration processes. \emph{Radio to optical polarimetry has since unveiled unique information about the magnetic field that cannot be obtained with continuum observations and light curves.} The observed polarization can be explained by intense particle acceleration processes in the jet emission region, such as shock \citep{Chen14,Chandra15,Zhang16}, magnetic reconnection \citep{zhang2018}, kink instability \citep{Zhang17,nalewajko2017}, or turbulence \citep{Laing80,marscher2014}, or global jet structure and evolution, such as a bending jet \citep{abdo2010} or helical magnetic field structure \citep{marscher2008,marscher2010}.

{\bf High-energy (X-ray to $\gamma$-ray energies) polarimetry will open up a new window and play a crucial role in exploring the extreme physics of high-energy radiation, neutrino production, and cosmic ray acceleration in AGN jets.} High-energy polarimetry can pinpoint the various radiation mechanisms in the high-energy blazar SED via the PF observations. Synergizing with the neutrino observation, high-energy polarimetry can probe the hadronic interactions and cosmic rays in AGN jets, which give rise to the electromagnetic counterparts of neutrinos. In particular, if ultra-high-energy cosmic rays (UHECRs) are accelerated in AGN jets, high-energy polarimetry provides exclusive information on the acceleration mechanism and physical conditions by unveiling the magnetic field structure and evolution. {\bf Supports in the next decade on high-energy missions with dedicated polarimetry capability and theoretical studies on high-energy polarization will add invaluable insights on the multi-messenger AGN jet studies.}

\vspace{\spbefore}
\section{Key Scientific Questions}
\label{key_question}
\vspace{\spafter}
\subsection{High-Energy Radiation and Neutrinos}
\vspace{\spafter}

\begin{wrapfigure}{r}{0.5\textwidth}
\vspace{-0.4in}
\begin{center}
\includegraphics[width=0.5\textwidth]{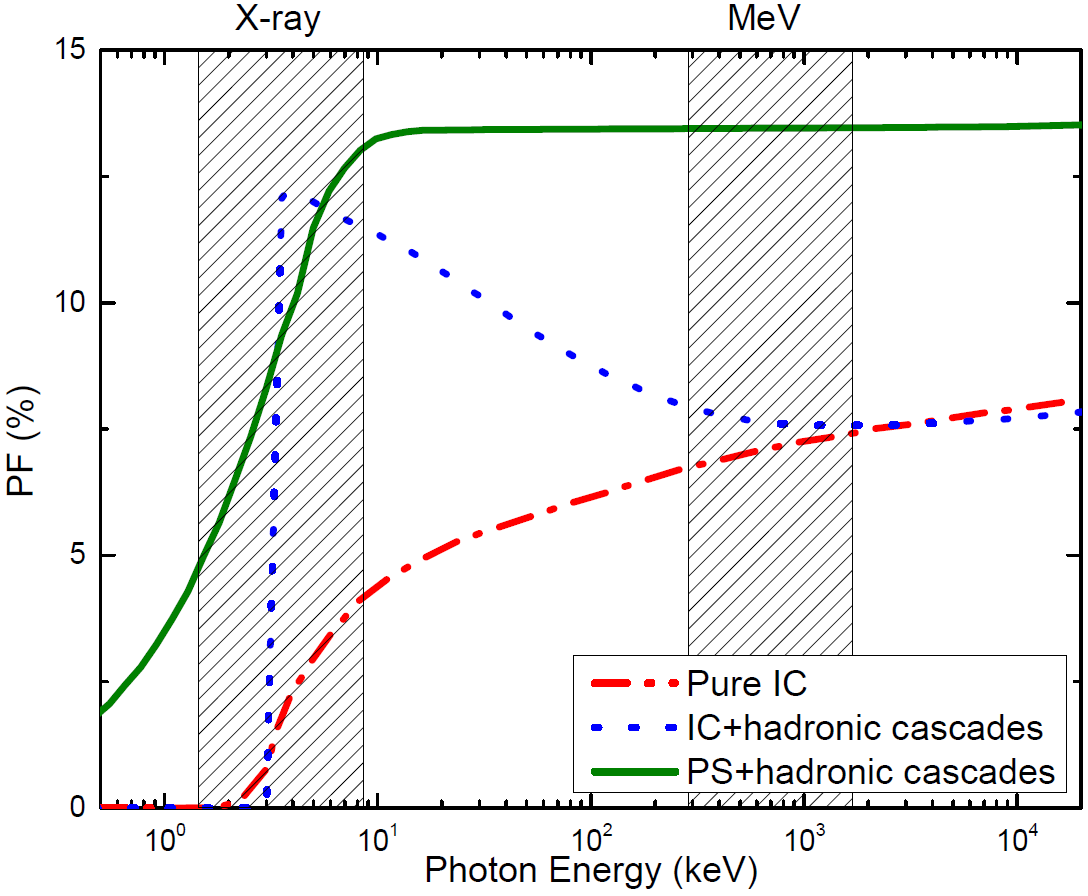}
\end{center}
\vspace{-0.8cm}
\captionsetup{font=footnotesize}
\caption{High-energy PF of TXS~0506+056 based on three different radiation mechanisms. The X-ray PF can probe the hadronic cascading pair synchrotron contribution, while the MeV $\gamma$-ray PF unambiguously distinguish IC and PS scenarios \citep{Zhang19}. \label{pic:specPol}}
\vspace{-0.2in}
\end{wrapfigure}

Blazar SEDs generally follow a double-hump spectral shape. While the low-energy hump is dominated by the synchrotron emission of non-thermal electrons,
there are two competing scenarios for the high-energy spectral component with distinct radiation mechanisms and underlying particles. The same non-thermal electrons that produce the low-energy hump can inverse Compton (IC) scatter low-energy photons to X-rays and $\gamma$-rays. The target photons for the IC scenario can be the primary electron synchrotron itself (synchrotron-self Compton, SSC) \citep{marscher85,maraschi92}, or external photon field (external Compton, EC) \citep{dermer92,sikora94} such as the thermal emission from accretion disk, broad line region, and dusty torus. The observed double-hump blazar SED suggests similar synchrotron and IC efficiency, inferring a magnetic field strength on the order of $0.1~\rm{G}$ in the blazar emission region \citep{bottcher2013}. Generally a pure leptonic IC model can successfully reproduce the blazar SED, but it may still include a subdominant hadronic component due to non-thermal protons, with neutrino counterparts \citep{Cerruti19,reimer18,Keivani18}. Current particle acceleration theories typically find that the acceleration mechanism that makes the non-thermal electrons can also accelerate non-thermal protons to very high energies \citep{sironi13,Guo16}. Interacting with the dense photon field in the emission region via photomeson processes, these protons can produce very high energy neutrinos and cascading pairs; the latter then results in secondary synchrotron emission from X-ray to MeV $\gamma$-rays \citep{bottcher2013,cerruti15,Petropoulou15}. 
If the blazar emission region has a strong magnetic field (typically $10-100~\rm{G}$), the efficient proton synchrotron ({\bf PS}) process can make X-rays and $\gamma$-rays \cite{mucke01,bottcher2013}. The high magnetic field in the PS scenario diminishes the IC efficiency of non-thermal electrons, so that the high-energy spectral component is mostly dominated by the PS and secondary synchrotron emission from hadronic cascading pairs \citep{Diltz15}. The observed GeV-TeV $\gamma$-rays imply the acceleration of UHECRs in the PS scenario \citep{bottcher2013,cerruti15,Zhang16b}. Both scenarios can successfully reproduce blazar SEDs. Although neutrinos can pinpoint the hadronic interactions, spectral modeling of the recent TXS~0506+056 event has found that the IC scenario with a subdominant hadronic component produces similar SEDs to those of the PS scenario \citep{Keivani18,Cerruti19}. Nonetheless, \emph{the drastically different magnetic field strengths in the two scenarios give rise to distinct radiation mechanisms, especially in the $\gamma$-rays.} We need additional observational constraints to distinguish between the different high-energy radiation mechanisms and probe the acceleration of cosmic rays and neutrino production.

\emph{High-energy polarimetry can disentangle the ambiguity in the radiation mechanism and underlying magnetic field strength.} This is because in the blazar emission environment, the PS scenario intrinsically predicts considerably higher PF than the IC scenario \citep{Bonometto70,Zhang13,Paliya18}. Figure \ref{pic:specPol} shows the PF in the high-energy component based on TXS~0506+056 model parameters \citep{Zhang19}. Assuming that the high-energy spectral component is produced co-spatially with the optical counterpart at $10\%$ PF, the PS scenario predicts a high PF at $\gtrsim 10\%$ in X-ray and MeV $\gamma$-ray bands, because of the synchrotron emission by either primary protons or secondary cascading pairs. A pure leptonic IC model predicts only $\sim 5\%$ PF in both energy bands, because the IC process generally makes negligible PF (EC) up to half of the synchrotron PF (SSC). 
In the case of an IC scenario with a subdominant hadronic contribution, 
the X-ray bands present a $\gtrsim 10\%$ PF similar to the PS scenario, due to the strong secondary pair synchrotron, but the MeV bands still show a much lower PF at $\sim 5\%$. Therefore, while the X-ray PF can probe the secondary pair synchrotron contribution complementary to the neutrino detection, \emph{MeV $\gamma$-ray polarization is key to unambiguously distinguish the IC and PS scenarios.}

\vspace{\spbefore}
\subsection{Particle Acceleration \label{acceleration}}
\vspace{\spafter}
\begin{wrapfigure}{r}{0.5\textwidth}
\vspace{-1.9cm}
\begin{center}
\includegraphics[width=0.48\textwidth]{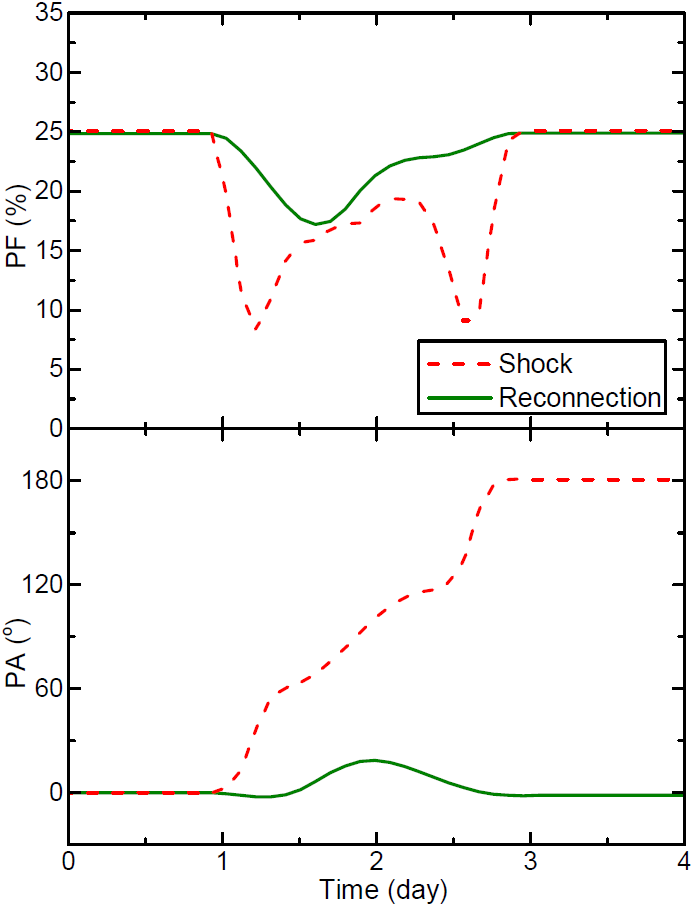}
\end{center}
\vspace{-0.8cm}
\captionsetup{font=footnotesize}
\caption{MeV $\gamma$-ray polarization variability of a PS dominated blazar, based on the shock and magnetic reconnection models. The shock model predicts a stronger polarization variability than the reconnection model \citep{Zhang16b}. \label{pic:tempPol}}
\vspace{-0.4cm}
\end{wrapfigure}
The fast  (on minutes timescales) $\gamma$-ray variability in blazars implies violent particle acceleration within very localized region(s) in jets \citep{ackermann2016}. Understanding the physical conditions and acceleration mechanisms therein is crucial to probe the cosmic ray acceleration and neutrino production in AGN jets. Theoretical studies have found that both shock and magnetic reconnection can efficiently dissipate plasma jet energy to accelerate non-thermal electrons and protons \citep{sironi13,Guo16}. Most importantly, \emph{the two mechanisms require very different magnetic energy composition in jets, and involve contrasting magnetic field evolution.} Shocks are efficient if the jet kinetic energy dominates over the magnetic energy \citep{Kirk00,Achterberg01}. They can convert bulk kinetic energy to accelerate non-thermal electrons and protons via the diffusive shock acceleration mechanism, and strongly alter the magnetic field structure \citep{marscher85,Laing80,sironi13}. On the other hand, magnetic reconnection usually requires the magnetic energy dominating over the kinetic energy. It generates magnetic plasmoids at the reconnection site, which are small chunks of plasma with high magnetic energy density \citep{giannios13,Sironi14,guo14}. This leads to efficient acceleration of non-thermal electrons and protons, and enhances the turbulence at the reconnection site. Nonetheless, although the two mechanisms involve very distinct physical conditions and evolution, both can reasonably explain the observed blazar SEDs and light curves \citep{marscher85,Spada01,Deng16,giannios09,romanova92,marscher2014,Baring17,christie19,tavecchio18}.

\emph{High-energy polarimetry can unveil the magnetic field evolution and disentangle the two particle acceleration mechanisms.} This is particularly important in exploring the cosmic ray acceleration, if the blazar high-energy spectrum is dominated by the PS scenario. Figure \ref{pic:tempPol} compares the temporal polarization signatures of the shock and magnetic reconnection under a PS blazar model, based on consistent numerical simulations of particle and magnetic field evolution \citep{Zhang16b}. The large change in the magnetic field morphology in the shocked plasma can lead to drastic MeV $\gamma$-ray polarization variation, such as a PA swing. However, magnetic reconnection only leads to more turbulent magnetic field, resulting in a moderate drop in the PF and minor variations in the PA. \emph{Future $\gamma$-ray polarimetry can examine the predictions of self-consistent first-principle simulations to disentangle the particle acceleration mechanisms in the blazar emission region, thus unveil the mystery of cosmic ray acceleration and neutrino production in AGN jets.}

\begin{figure}
\vspace{-0.5cm}
\includegraphics[trim=1 35 0 70, clip=true, scale=0.31]{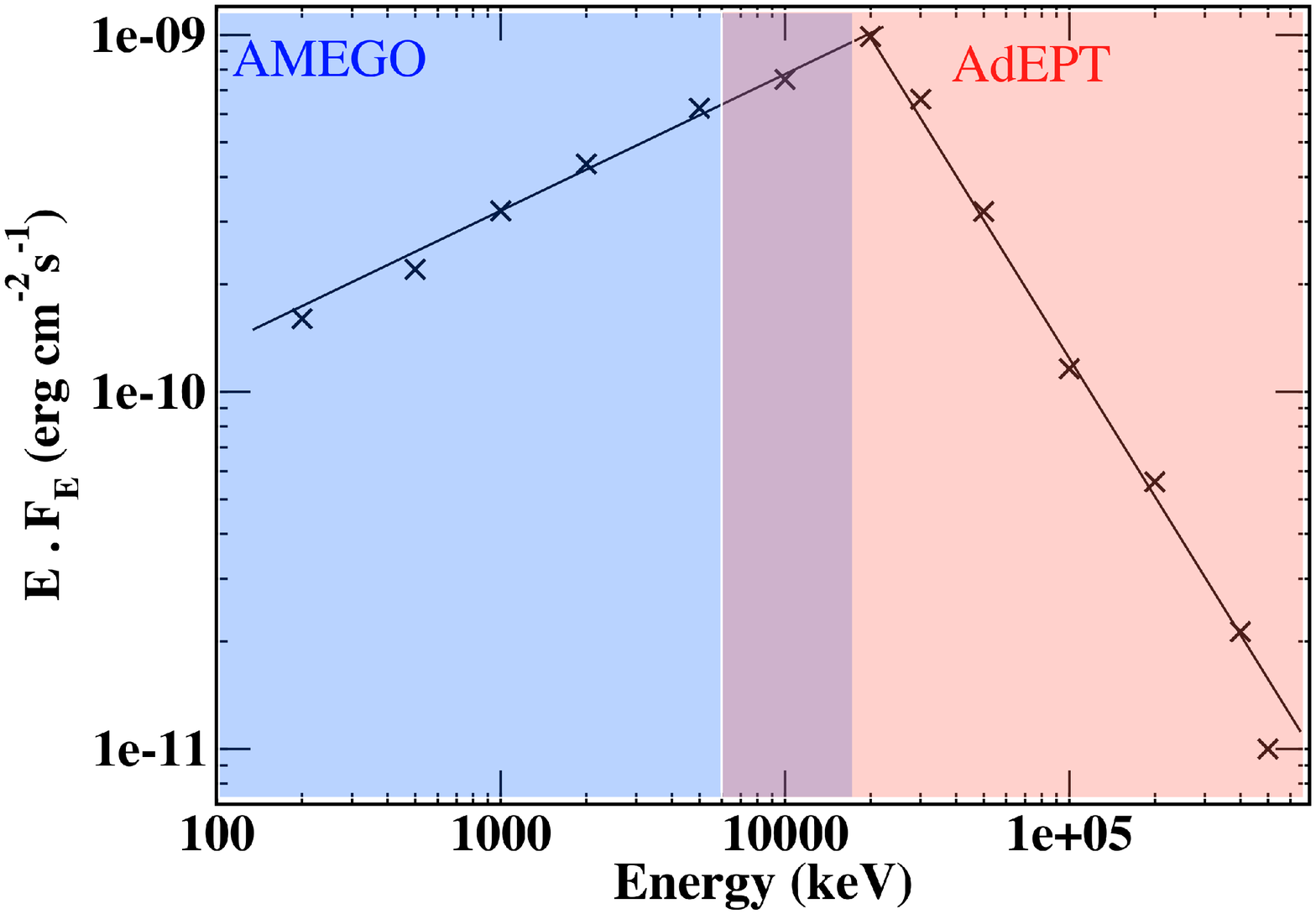}
\includegraphics[scale=0.3,trim=0 0 0 3,clip=true]{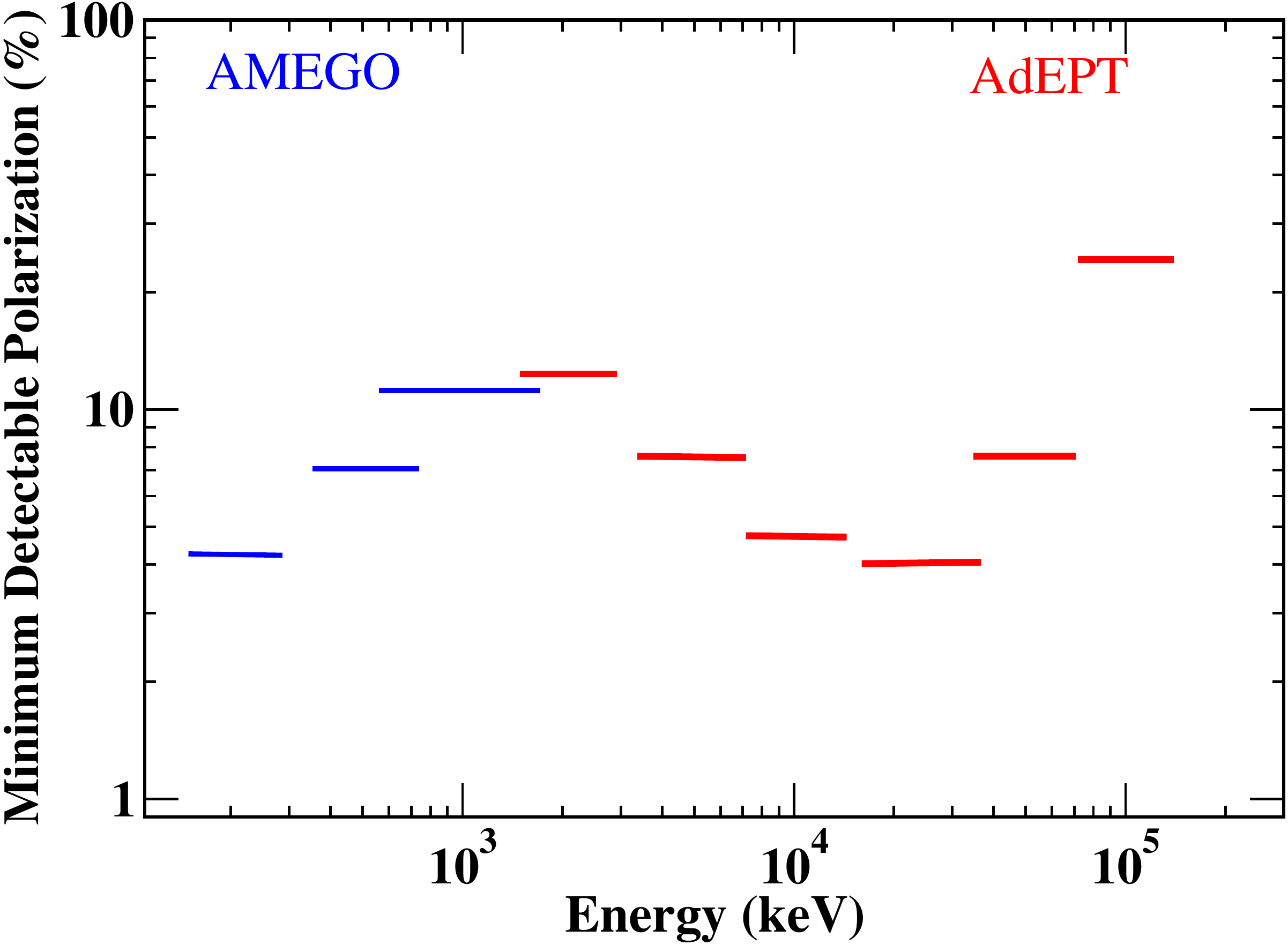}
\captionsetup{font=footnotesize}
\caption{ {\it Left:} Spectral energy distribution blazar 3C~279 during a typical bright state \citep{hayashida2015} used for the simulations of polarization sensitivity for AMEGO (blue) and AdEPT (red).  
{\it Right:} Predicted minimum detectable polarization (MDP) for AMEGO (blue) and AdEPT (red) for a 20-day on-source integration time.}
\label{fig_sim}
\vspace{-0.5cm}
\end{figure}

\vspace{\spbefore}
\section{Required Instrumental and Theoretical Advances}
\vspace{\spafter}
\subsection{High-energy polarimetry - a window about to open}
\vspace{\spafter}

Addressing the science goals above requires multi-wavelength and multi-messenger observations with wide field of view, good energy resolution, and high sensitivity.  
In particular, future high-energy missions with dedicated polarimetry capability will be optimal. Several missions, e.g. {\bf XPP}, {\bf IXPE}\footnote{\url{https://ixpe.msfc.nasa.gov/}} \citep{ixpe}, {\bf e-ASTROGAM}\footnote{\url{http://eastrogam.iaps.inaf.it/science.html}} \citep{eastrogam}, {\bf AMEGO}\footnote{\url{https://asd.gsfc.nasa.gov/amego/}} \citep{amego2019}, and  {\bf AdEPT} \cite{adept}, are planned in the next decade 
with great polarization sensitivity over the energy range from 2 keV to 500~MeV.

Figure \ref{fig_sim} illustrates the polarization detection capabilities  in the keV and MeV energy range. 
The polarization detection capabilities of the future MeV missions AMEGO and AdEPT  were calculated on the 
assumption of  blazar, 3C~279, with a degree of polarization of 20$\%$ and broken power-law energy flux spectrum 
during a typical bright state, energy-flux at 100~MeV i.e.\ $F_{100~MeV}$ = 10$^{-10}~\rm{erg~cm^{-2}~s^{-1}}$ \citep{hayashida2015}.  
The AMEGO minimum detectable polarization (MDP) in the 100~keV to 5~MeV energy range was simulated using the MEGALIB tool\footnote{http://megalibtoolkit.com/home.html}.  The AdEPT MDP in the 5-200~MeV energy range was calculated using the instrument parameters given by \citet{adept}. {\it Simulations predict that we will be able to achieve  MDP as low as 5$\%$ for both missions}, making the detection of high-energy polarization signals from many AGN very promising.

\begin{wrapfigure}{r}{0.42\textwidth}
\vspace{-0.65cm}
\includegraphics[scale=0.27,trim=10 0 0 60,clip=true]{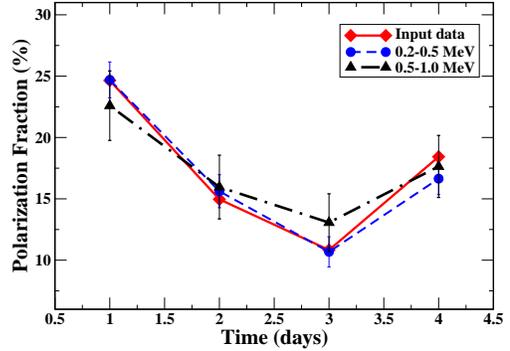}
\vspace{-1.1cm}
\captionsetup{font=footnotesize}
\caption{ Predicted changes in PF during a typical flare, energy flux~=~10$^{-9}~\rm{erg~cm^{-2}~s^{-1}}$, at 0.2-0.5~MeV (blue) and 0.5-1.0~MeV (black) energies. The red diamonds are the input data.  }
\label{fig_pol_curve}
\vspace{-0.65cm}
\end{wrapfigure}
Simulations imply that AMEGO will be able to detect polarization signals from blazars  with a flux above $F_{100~MeV}$. In four years of {\it Fermi}/LAT observations 55 blazars were detected with a peak energy-flux above $F_{100~MeV}$ \cite{3fgl}, {\bf extrapolating this indicates that $\sim$68 blazars would be detected during the 5-year  mission}.  These calculations also predict that during flaring states of bright blazars, i.e.\ an increase in flux by an order of magnitude, we will be able to not only detect more sources but also probe temporal polarization variations.
Figure \ref{fig_pol_curve} displays the predicted  variations in PF 
on day scale during a typical flaring state of the blazar 3C~279 (energy flux $\geq$10$^{-9}~\rm{erg~cm^{-2}~s^{-1}}$) at 0.2-0.5~MeV (in blue) and 0.5-1.0~MeV (in black) energies (the input data is shown in red). The simulated temporal polarization sensitivity is very appealing  to diagnose the cosmic ray acceleration mechanisms as shown in \S\ref{acceleration}.

\vspace{\spbefore}
\subsection{Theoretical advances}
\vspace{\spafter}
In order to leverage the multi-messenger observation and multi-wavelength polarimetry in the next decade to study the AGN jet physics and potential cosmic ray and neutrino production, we need to 
achieve a through physical understanding of the plasma jet dynamics, particle acceleration, and radiation processes. The highly variable and complex nature of AGN jets prevents the use of simple analytical models with many free parameters, as they will diminish the model prediction power. Instead, we need 
comprehensive models based on first principles theories and numerical simulations to minimize the parameter freedom and reveal the dynamical fluid and particle co-evolution, as well as the radiative transfer and feedback.

Full particle-in-cell (PIC) simulations have been successful in studying the particle acceleration processes during shocks and magnetic reconnection events \citep{Spitkovsky08,Sironi09,Nishikawa05,Sironi14,guo14,Werner18}. This approach can self-consistently evolve both fluid and particle physical quantities, at the cost of high computational power. However, PIC can only simulate evolution on particle kinetic scale, which is much smaller than the blazar emission region \citep{Sironi14,Guo16,Werner18}. 
Several multi-scale approaches have been developed, where some kinetic effects are removed to reduce the computational cost and push simulations to larger scales \citep[e.g.,][]{Achterberg01,Drake19,Li18,Bai15}. For example, some models treat thermal electrons as a fluid and only retain non-thermal electrons and protons. Much further studies are needed to determine the necessary kinetic physics for understanding the particle acceleration processes. \emph{Supports on development and applications of multi-scale numerical simulations and high-performance computing in astrophysics will extend our understanding to multi-messenger astrophysics.}

The ultra-relativistic electrons and cosmic rays in the blazar emission region mostly lose energy through radiation. Therefore, we need to study radiation transfer for both electromagnetic and neutrino signatures, as well as radiative feedback on non-thermal particles. A couple of radiation transfer codes are advancing in this direction, using Monte-Carlo and/or ray-tracing methods \citep{Chen14,Zhang14,marscher2014}. Given that the blazar emission region can be highly inhomogeneous and fast evolving, the radiation transfer simulations need to consider anisotropy and spatial inhomogeneity effects and be fully coupled with fluid and particle dynamics. \emph{Supports on numerical methods on first-principle integrated radiative transfer simulations will play a central role in the multi-messenger astronomy.}

\clearpage


\begin{thebibliography}{}
\expandafter\ifx\csname natexlab\endcsname\relax\def\natexlab#1{#1}\fi
\providecommand{\url}[1]{\href{#1}{#1}}
\providecommand{\dodoi}[1]{doi:~\href{http://doi.org/#1}{\nolinkurl{#1}}}
\providecommand{\doeprint}[1]{\href{http://ascl.net/#1}{\nolinkurl{http://ascl.net/#1}}}
\providecommand{\doarXiv}[1]{\href{https://arxiv.org/abs/#1}{\nolinkurl{https://arxiv.org/abs/#1}}}

\bibitem[{{Abdo} {et~al.}(2010){Abdo}, {Ackermann}, {Ajello}, {Axelsson},
  {Baldini}, {Ballet}, {Barbiellini}, {Bastieri}, {Baughman}, {Bechtol}, \&
  et~al.}]{abdo2010}
{Abdo}, A.~A., {Ackermann}, M., {Ajello}, M., {et~al.} 2010, \nat, 463, 919,
  \dodoi{10.1038/nature08841}

\bibitem[{{Abdo} {et~al.}(2011){Abdo}, {Ackermann}, {Ajello}, {Allafort},
  {Baldini}, {Ballet}, {Barbiellini}, {Bastieri}, {Bellazzini}, {Berenji},
  {Blandford}, {Bloom}, {Bonamente}, {Borgland}, {Bouvier}, {Bregeon},
  {Brigida}, {Bruel}, {Buehler}, {Buson}, {Caliandro}, {Cameron}, {Caraveo},
  {Casandjian}, {Cavazzuti}, {Cecchi}, {Charles}, {Chekhtman}, {Cheung},
  {Chiang}, {Ciprini}, {Claus}, {Conrad}, {Cutini}, {D'Ammando}, {de Angelis},
  {de Palma}, {Dermer}, {Digel}, {Silva}, {Drell}, {Dubois}, {Dumora},
  {Escande}, {Favuzzi}, {Fegan}, {Ferrara}, {Fortin}, {Fukazawa}, {Fusco},
  {Gargano}, {Gasparrini}, {Gehrels}, {Germani}, {Giglietto}, {Giommi},
  {Giordano}, {Giroletti}, {Glanzman}, {Godfrey}, {Grenier}, {Grove},
  {Guiriec}, {Hadasch}, {Hayashida}, {Hays}, {Horan}, {Itoh},
  {J{\'o}hannesson}, {Johnson}, {Kamae}, {Katagiri}, {Kataoka},
  {Kn{\"o}dlseder}, {Kuss}, {Lande}, {Larsson}, {Latronico}, {Lee}, {Longo},
  {Loparco}, {Lott}, {Lovellette}, {Lubrano}, {Madejski}, {Makeev},
  {Mazziotta}, {McConville}, {McEnery}, {Michelson}, {Mitthumsiri}, {Mizuno},
  {Moiseev}, {Monte}, {Monzani}, {Morselli}, {Moskalenko}, {Murgia},
  {Naumann-Godo}, {Nishino}, {Nolan}, {Norris}, {Nuss}, {Ohsugi}, {Okumura},
  {Orlando}, {Ormes}, {Paneque}, {Pelassa}, {Pesce-Rollins}, {Pierbattista},
  {Piron}, {Porter}, {Rain{\`o}}, {Rando}, {Razzaque}, {Reimer}, {Reimer},
  {Ritz}, {Roth}, {Sadrozinski}, {Sanchez}, {Scargle}, {Schalk}, {Sgr{\`o}},
  {Siskind}, {Smith}, {Spandre}, {Spinelli}, {Strickman}, {Takahashi},
  {Takahashi}, {Tanaka}, {Tanaka}, {Thayer}, {Thayer}, {Thompson}, {Tibaldo},
  {Torres}, {Tosti}, {Tramacere}, {Troja}, {Vandenbroucke}, {Vasileiou},
  {Vianello}, {Vilchez}, {Vitale}, {Waite}, {Wang}, {Winer}, {Wood}, {Yang}, \&
  {Ziegler}}]{abdo2011}
---. 2011, \apjl, 733, L26, \dodoi{10.1088/2041-8205/733/2/L26}

\bibitem[{{Acero} {et~al.}(2015){Acero}, {Ackermann}, {Ajello}, {Albert},
  {Atwood}, {Axelsson}, {Baldini}, {Ballet}, {Barbiellini}, {Bastieri},
  {Belfiore}, {Bellazzini}, {Bissaldi}, {Blandford}, {Bloom}, {Bogart},
  {Bonino}, {Bottacini}, {Bregeon}, {Britto}, {Bruel}, {Buehler}, {Burnett},
  {Buson}, {Caliandro}, {Cameron}, {Caputo}, {Caragiulo}, {Caraveo},
  {Casandjian}, {Cavazzuti}, {Charles}, {Chaves}, {Chekhtman}, {Cheung},
  {Chiang}, {Chiaro}, {Ciprini}, {Claus}, {Cohen-Tanugi}, {Cominsky}, {Conrad},
  {Cutini}, {D'Ammando}, {de Angelis}, {DeKlotz}, {de Palma}, {Desiante},
  {Digel}, {Di Venere}, {Drell}, {Dubois}, {Dumora}, {Favuzzi}, {Fegan},
  {Ferrara}, {Finke}, {Franckowiak}, {Fukazawa}, {Funk}, {Fusco}, {Gargano},
  {Gasparrini}, {Giebels}, {Giglietto}, {Giommi}, {Giordano}, {Giroletti},
  {Glanzman}, {Godfrey}, {Grenier}, {Grondin}, {Grove}, {Guillemot}, {Guiriec},
  {Hadasch}, {Harding}, {Hays}, {Hewitt}, {Hill}, {Horan}, {Iafrate}, {Jogler},
  {J{\'o}hannesson}, {Johnson}, {Johnson}, {Johnson}, {Johnson}, {Kamae},
  {Kataoka}, {Katsuta}, {Kuss}, {La Mura}, {Landriu}, {Larsson}, {Latronico},
  {Lemoine-Goumard}, {Li}, {Li}, {Longo}, {Loparco}, {Lott}, {Lovellette},
  {Lubrano}, {Madejski}, {Massaro}, {Mayer}, {Mazziotta}, {McEnery},
  {Michelson}, {Mirabal}, {Mizuno}, {Moiseev}, {Mongelli}, {Monzani},
  {Morselli}, {Moskalenko}, {Murgia}, {Nuss}, {Ohno}, {Ohsugi}, {Omodei},
  {Orienti}, {Orlando}, {Ormes}, {Paneque}, {Panetta}, {Perkins},
  {Pesce-Rollins}, {Piron}, {Pivato}, {Porter}, {Racusin}, {Rando}, {Razzano},
  {Razzaque}, {Reimer}, {Reimer}, {Reposeur}, {Rochester}, {Romani},
  {Salvetti}, {S{\'a}nchez-Conde}, {Saz Parkinson}, {Schulz}, {Siskind},
  {Smith}, {Spada}, {Spandre}, {Spinelli}, {Stephens}, {Strong}, {Suson},
  {Takahashi}, {Takahashi}, {Tanaka}, {Thayer}, {Thayer}, {Thompson},
  {Tibaldo}, {Tibolla}, {Torres}, {Torresi}, {Tosti}, {Troja}, {Van Klaveren},
  {Vianello}, {Winer}, {Wood}, {Wood}, {Zimmer}, \& {Fermi-LAT
  Collaboration}}]{3fgl}
{Acero}, F., {Ackermann}, M., {Ajello}, M., {et~al.} 2015, ApJS, 218, 23,
  \dodoi{10.1088/0067-0049/218/2/23}

\bibitem[{{Achterberg} {et~al.}(2001){Achterberg}, {Gallant}, {Kirk}, \&
  {Guthmann}}]{Achterberg01}
{Achterberg}, A., {Gallant}, Y.~A., {Kirk}, J.~G., \& {Guthmann}, A.~W. 2001,
  \mnras, 328, 393, \dodoi{10.1046/j.1365-8711.2001.04851.x}

\bibitem[{{Ackermann} {et~al.}(2015){Ackermann}, {Ajello}, {Atwood}, {Baldini},
  {Ballet}, {Barbiellini}, {Bastieri}, {Becerra Gonzalez}, {Bellazzini},
  {Bissaldi}, {Blandford}, {Bloom}, {Bonino}, {Bottacini}, {Brandt}, {Bregeon},
  {Britto}, {Bruel}, {Buehler}, {Buson}, {Caliandro}, {Cameron}, {Caragiulo},
  {Caraveo}, {Carpenter}, {Casandjian}, {Cavazzuti}, {Cecchi}, {Charles},
  {Chekhtman}, {Cheung}, {Chiang}, {Chiaro}, {Ciprini}, {Claus},
  {Cohen-Tanugi}, {Cominsky}, {Conrad}, {Cutini}, {D'Abrusco}, {D'Ammando}, {de
  Angelis}, {Desiante}, {Digel}, {Di Venere}, {Drell}, {Favuzzi}, {Fegan},
  {Ferrara}, {Finke}, {Focke}, {Franckowiak}, {Fuhrmann}, {Fukazawa},
  {Furniss}, {Fusco}, {Gargano}, {Gasparrini}, {Giglietto}, {Giommi},
  {Giordano}, {Giroletti}, {Glanzman}, {Godfrey}, {Grenier}, {Grove},
  {Guiriec}, {Hewitt}, {Hill}, {Horan}, {Itoh}, {J{\'o}hannesson}, {Johnson},
  {Johnson}, {Kataoka}, {Kawano}, {Krauss}, {Kuss}, {La Mura}, {Larsson},
  {Latronico}, {Leto}, {Li}, {Li}, {Longo}, {Loparco}, {Lott}, {Lovellette},
  {Lubrano}, {Madejski}, {Mayer}, {Mazziotta}, {McEnery}, {Michelson},
  {Mizuno}, {Moiseev}, {Monzani}, {Morselli}, {Moskalenko}, {Murgia}, {Nuss},
  {Ohno}, {Ohsugi}, {Ojha}, {Omodei}, {Orienti}, {Orlando}, {Paggi}, {Paneque},
  {Perkins}, {Pesce-Rollins}, {Piron}, {Pivato}, {Porter}, {Rain{\`o}},
  {Rando}, {Razzano}, {Razzaque}, {Reimer}, {Reimer}, {Romani}, {Salvetti},
  {Schaal}, {Schinzel}, {Schulz}, {Sgr{\`o}}, {Siskind}, {Sokolovsky}, {Spada},
  {Spandre}, {Spinelli}, {Stawarz}, {Suson}, {Takahashi}, {Takahashi},
  {Tanaka}, {Thayer}, {Thayer}, {Tibaldo}, {Torres}, {Torresi}, {Tosti},
  {Troja}, {Uchiyama}, {Vianello}, {Winer}, {Wood}, \& {Zimmer}}]{3lac2015}
{Ackermann}, M., {Ajello}, M., {Atwood}, W.~B., {et~al.} 2015, \apj, 810, 14,
  \dodoi{10.1088/0004-637X/810/1/14}

\bibitem[{{Ackermann} {et~al.}(2016){Ackermann}, {Anantua}, {Asano}, {Baldini},
  {Barbiellini}, {Bastieri}, {Becerra Gonzalez}, {Bellazzini}, {Bissaldi},
  {Blandford}, {Bloom}, {Bonino}, {Bottacini}, {Bruel}, {Buehler}, {Caliandro},
  {Cameron}, {Caragiulo}, {Caraveo}, {Cavazzuti}, {Cecchi}, {Cheung}, {Chiang},
  {Chiaro}, {Ciprini}, {Cohen-Tanugi}, {Costanza}, {Cutini}, {D'Ammando}, {de
  Palma}, {Desiante}, {Digel}, {Di Lalla}, {Di Mauro}, {Di Venere}, {Drell},
  {Favuzzi}, {Fegan}, {Ferrara}, {Fukazawa}, {Funk}, {Fusco}, {Gargano},
  {Gasparrini}, {Giglietto}, {Giordano}, {Giroletti}, {Grenier}, {Guillemot},
  {Guiriec}, {Hayashida}, {Hays}, {Horan}, {J{\'o}hannesson}, {Kensei},
  {Kocevski}, {Kuss}, {La Mura}, {Larsson}, {Latronico}, {Li}, {Longo},
  {Loparco}, {Lott}, {Lovellette}, {Lubrano}, {Madejski}, {Magill}, {Maldera},
  {Manfreda}, {Mayer}, {Mazziotta}, {Michelson}, {Mirabal}, {Mizuno},
  {Monzani}, {Morselli}, {Moskalenko}, {Nalewajko}, {Negro}, {Nuss}, {Ohsugi},
  {Orlando}, {Paneque}, {Perkins}, {Pesce-Rollins}, {Piron}, {Pivato},
  {Porter}, {Principe}, {Rando}, {Razzano}, {Razzaque}, {Reimer}, {Scargle},
  {Sgr{\`o}}, {Sikora}, {Simone}, {Siskind}, {Spada}, {Spinelli}, {Stawarz},
  {Thayer}, {Thompson}, {Torres}, {Troja}, {Uchiyama}, {Yuan}, \&
  {Zimmer}}]{ackermann2016}
{Ackermann}, M., {Anantua}, R., {Asano}, K., {et~al.} 2016, \apjl, 824, L20,
  \dodoi{10.3847/2041-8205/824/2/L20}

\bibitem[{{Agudo} {et~al.}(2011){Agudo}, {Jorstad}, {Marscher}, {Larionov},
  {G{\'o}mez}, {L{\"a}hteenm{\"a}ki}, {Gurwell}, {Smith}, {Wiesemeyer}, {Thum},
  {Heidt}, {Blinov}, {D'Arcangelo}, {Hagen-Thorn}, {Morozova}, {Nieppola},
  {Roca-Sogorb}, {Schmidt}, {Taylor}, {Tornikoski}, \& {Troitsky}}]{agudo2011}
{Agudo}, I., {Jorstad}, S.~G., {Marscher}, A.~P., {et~al.} 2011, \apjl, 726,
  L13, \dodoi{10.1088/2041-8205/726/1/L13}

\bibitem[{{Bai} {et~al.}(2015){Bai}, {Caprioli}, {Sironi}, \&
  {Spitkovsky}}]{Bai15}
{Bai}, X.-N., {Caprioli}, D., {Sironi}, L., \& {Spitkovsky}, A. 2015, \apj,
  809, 55, \dodoi{10.1088/0004-637X/809/1/55}

\bibitem[{{Baring} {et~al.}(2017){Baring}, {B{\"o}ttcher}, \&
  {Summerlin}}]{Baring17}
{Baring}, M.~G., {B{\"o}ttcher}, M., \& {Summerlin}, E.~J. 2017, \mnras, 464,
  4875, \dodoi{10.1093/mnras/stw2344}

\bibitem[{{Bhatta} {et~al.}(2015){Bhatta}, {Goyal}, {Ostrowski}, {Stawarz},
  {Akitaya}, {Arkharov}, {Bachev}, {Ben{\'{\i}}tez}, {Borman}, {Carosati},
  {Cason}, {Damljanovic}, {Dhalla}, {Frasca}, {Hu}, {Itoh}, {Jorstad},
  {Jableka}, {Kawabata}, {Klimanov}, {Kurtanidze}, {Larionov}, {Laurence},
  {Leto}, {Markowitz}, {Marscher}, {Moody}, {Moritani}, {Ohlert}, {Di Paola},
  {Raiteri}, {Rizzi}, {Sadun}, {Sasada}, {Sergeev}, {Strigachev}, {Takaki},
  {Troitsky}, {Ui}, {Villata}, {Vince}, {Webb}, {Yoshida}, {Zola}, \&
  {Hiriart}}]{bhatta2015}
{Bhatta}, G., {Goyal}, A., {Ostrowski}, M., {et~al.} 2015, \apjl, 809, L27,
  \dodoi{10.1088/2041-8205/809/2/L27}

\bibitem[{{Blinov} {et~al.}(2016){Blinov}, {Pavlidou}, {Papadakis}, {Hovatta},
  {Pearson}, {Liodakis}, {Panopoulou}, {Angelakis}, {Balokovi{\'c}}, {Das},
  {Khodade}, {Kiehlmann}, {King}, {Kus}, {Kylafis}, {Mahabal}, {Marecki},
  {Modi}, {Myserlis}, {Paleologou}, {Papamastorakis}, {Pazderska}, {Pazderski},
  {Rajarshi}, {Ramaprakash}, {Readhead}, {Reig}, {Tassis}, \&
  {Zensus}}]{blinov16}
{Blinov}, D., {Pavlidou}, V., {Papadakis}, I.~E., {et~al.} 2016, \mnras, 457,
  2252, \dodoi{10.1093/mnras/stw158}

\bibitem[{{Blinov} {et~al.}(2018){Blinov}, {Pavlidou}, {Papadakis},
  {Kiehlmann}, {Liodakis}, {Panopoulou}, {Angelakis}, {Balokovi{\'c}},
  {Hovatta}, {King}, {Kus}, {Kylafis}, {Mahabal}, {Maharana}, {Myserlis},
  {Paleologou}, {Papamastorakis}, {Pazderski}, {Pearson}, {Ramaprakash},
  {Readhead}, {Reig}, {Tassis}, \& {Zensus}}]{blinov2018}
{Blinov}, D., {Pavlidou}, V., {Papadakis}, I., {et~al.} 2018, \mnras, 474,
  1296, \dodoi{10.1093/mnras/stx2786}

\bibitem[{{Bondi} {et~al.}(2004){Bondi}, {March{\~a}}, {Polatidis},
  {Dallacasa}, {Stanghellini}, \& {Ant{\'o}n}}]{bondi2004}
{Bondi}, M., {March{\~a}}, M.~J.~M., {Polatidis}, A., {et~al.} 2004, \mnras,
  352, 112, \dodoi{10.1111/j.1365-2966.2004.07903.x}

\bibitem[{{Bonometto} {et~al.}(1970){Bonometto}, {Cazzola}, \&
  {Saggion}}]{Bonometto70}
{Bonometto}, S., {Cazzola}, P., \& {Saggion}, A. 1970, \aap, 7, 292

\bibitem[{{B{\"o}ttcher} {et~al.}(2013){B{\"o}ttcher}, {Reimer}, {Sweeney}, \&
  {Prakash}}]{bottcher2013}
{B{\"o}ttcher}, M., {Reimer}, A., {Sweeney}, K., \& {Prakash}, A. 2013, \apj,
  768, 54, \dodoi{10.1088/0004-637X/768/1/54}

\bibitem[{{Casadio} {et~al.}(2019){Casadio}, {Marscher}, {Jorstad}, {Blinov},
  {MacDonald}, {Krichbaum}, {Boccardi}, {Traianou}, {G{\'o}mez}, {Agudo},
  {Sohn}, {Bremer}, {Hodgson}, {Kallunki}, {Kim}, {Williamson}, \&
  {Zensus}}]{casadio2019}
{Casadio}, C., {Marscher}, A.~P., {Jorstad}, S.~G., {et~al.} 2019, \aap, 622,
  A158, \dodoi{10.1051/0004-6361/201834519}

\bibitem[{{Cerruti} {et~al.}(2019){Cerruti}, {Zech}, {Boisson}, {Emery},
  {Inoue}, \& {Lenain}}]{Cerruti19}
{Cerruti}, M., {Zech}, A., {Boisson}, C., {et~al.} 2019, \mnras, 483, L12,
  \dodoi{10.1093/mnrasl/sly210}

\bibitem[{{Cerruti} {et~al.}(2015){Cerruti}, {Zech}, {Boisson}, \&
  {Inoue}}]{cerruti15}
{Cerruti}, M., {Zech}, A., {Boisson}, C., \& {Inoue}, S. 2015, \mnras, 448,
  910, \dodoi{10.1093/mnras/stu2691}

\bibitem[{{Chandra} {et~al.}(2015){Chandra}, {Zhang}, {Kushwaha}, {Singh},
  {Bottcher}, {Kaur}, \& {Baliyan}}]{Chandra15}
{Chandra}, S., {Zhang}, H., {Kushwaha}, P., {et~al.} 2015, \apj, 809, 130,
  \dodoi{10.1088/0004-637X/809/2/130}

\bibitem[{{Chen} {et~al.}(2014){Chen}, {Chatterjee}, {Zhang}, {Pohl},
  {Fossati}, {B{\"o}ttcher}, {Bailyn}, {Bonning}, {Buxton}, {Coppi}, {Isler},
  {Maraschi}, \& {Urry}}]{Chen14}
{Chen}, X., {Chatterjee}, R., {Zhang}, H., {et~al.} 2014, \mnras, 441, 2188,
  \dodoi{10.1093/mnras/stu713}

\bibitem[{{Christie} {et~al.}(2019){Christie}, {Petropoulou}, {Sironi}, \&
  {Giannios}}]{christie19}
{Christie}, I.~M., {Petropoulou}, M., {Sironi}, L., \& {Giannios}, D. 2019,
  \mnras, 482, 65, \dodoi{10.1093/mnras/sty2636}

\bibitem[{{de Angelis} {et~al.}(2018){de Angelis}, {Tatischeff}, {Grenier},
  {McEnery}, {Mallamaci}, {Tavani}, {Oberlack}, {Hanlon}, {Walter}, {Argan}, \&
  et~al.}]{eastrogam}
{de Angelis}, A., {Tatischeff}, V., {Grenier}, I.~A., {et~al.} 2018, Journal of
  High Energy Astrophysics, 19, 1, \dodoi{10.1016/j.jheap.2018.07.001}

\bibitem[{{Deng} {et~al.}(2016){Deng}, {Zhang}, {Zhang}, \& {Li}}]{Deng16}
{Deng}, W., {Zhang}, H., {Zhang}, B., \& {Li}, H. 2016, \apjl, 821, L12,
  \dodoi{10.3847/2041-8205/821/1/L12}

\bibitem[{{Dermer} {et~al.}(1992){Dermer}, {Schlickeiser}, \&
  {Mastichiadis}}]{dermer92}
{Dermer}, C.~D., {Schlickeiser}, R., \& {Mastichiadis}, A. 1992, \aap, 256, L27

\bibitem[{{Diltz} {et~al.}(2015){Diltz}, {B{\"o}ttcher}, \&
  {Fossati}}]{Diltz15}
{Diltz}, C., {B{\"o}ttcher}, M., \& {Fossati}, G. 2015, \apj, 802, 133,
  \dodoi{10.1088/0004-637X/802/2/133}

\bibitem[{{Drake} {et~al.}(2019){Drake}, {Arnold}, {Swisdak}, \&
  {Dahlin}}]{Drake19}
{Drake}, J.~F., {Arnold}, H., {Swisdak}, M., \& {Dahlin}, J.~T. 2019, Physics
  of Plasmas, 26, 012901, \dodoi{10.1063/1.5058140}

\bibitem[{{Giannios}(2013)}]{giannios13}
{Giannios}, D. 2013, \mnras, 431, 355, \dodoi{10.1093/mnras/stt167}

\bibitem[{{Giannios} {et~al.}(2009){Giannios}, {Uzdensky}, \&
  {Begelman}}]{giannios09}
{Giannios}, D., {Uzdensky}, D.~A., \& {Begelman}, M.~C. 2009, \mnras, 395, L29,
  \dodoi{10.1111/j.1745-3933.2009.00635.x}

\bibitem[{{Guo} {et~al.}(2014){Guo}, {Li}, {Daughton}, \& {Liu}}]{guo14}
{Guo}, F., {Li}, H., {Daughton}, W., \& {Liu}, Y.-H. 2014, Physical Review
  Letters, 113, 155005, \dodoi{10.1103/PhysRevLett.113.155005}

\bibitem[{{Guo} {et~al.}(2016){Guo}, {Li}, {Li}, {Daughton}, {Zhang},
  {Lloyd-Ronning}, {Liu}, {Zhang}, \& {Deng}}]{Guo16}
{Guo}, F., {Li}, X., {Li}, H., {et~al.} 2016, \apjl, 818, L9,
  \dodoi{10.3847/2041-8205/818/1/L9}

\bibitem[{{Hayashida} {et~al.}(2015){Hayashida}, {Nalewajko}, {Madejski},
  {Sikora}, {Itoh}, {Ajello}, {Blandford}, {Buson}, {Chiang}, {Fukazawa},
  {Furniss}, {Urry}, {Hasan}, {Harrison}, {Alexander}, {Balokovi{\'c}},
  {Barret}, {Boggs}, {Christensen}, {Craig}, {Forster}, {Giommi},
  {Grefenstette}, {Hailey}, {Hornstrup}, {Kitaguchi}, {Koglin}, {Madsen},
  {Mao}, {Miyasaka}, {Mori}, {Perri}, {Pivovaroff}, {Puccetti}, {Rana},
  {Stern}, {Tagliaferri}, {Westergaard}, {Zhang}, {Zoglauer}, {Gurwell},
  {Uemura}, {Akitaya}, {Kawabata}, {Kawaguchi}, {Kanda}, {Moritani}, {Takaki},
  {Ui}, {Yoshida}, {Agarwal}, \& {Gupta}}]{hayashida2015}
{Hayashida}, M., {Nalewajko}, K., {Madejski}, G.~M., {et~al.} 2015, \apj, 807,
  79, \dodoi{10.1088/0004-637X/807/1/79}

\bibitem[{{Hunter} {et~al.}(2014){Hunter}, {Bloser}, {Depaola}, {Dion},
  {DeNolfo}, {Hanu}, {Iparraguirre}, {Legere}, {Longo}, {McConnell}, {Nowicki},
  {Ryan}, {Son}, \& {Stecker}}]{adept}
{Hunter}, S.~D., {Bloser}, P.~F., {Depaola}, G.~O., {et~al.} 2014,
  Astroparticle Physics, 59, 18, \dodoi{10.1016/j.astropartphys.2014.04.002}

\bibitem[{{IceCube Collaboration} {et~al.}(2018){IceCube Collaboration},
  {Aartsen}, {Ackermann}, {Adams}, {Aguilar}, {Ahlers}, {Ahrens}, {Al Samarai},
  {Altmann}, {Andeen}, \& et~al.}]{icecube2018}
{IceCube Collaboration}, {Aartsen}, M.~G., {Ackermann}, M., {et~al.} 2018,
  Science, 361, eaat1378, \dodoi{10.1126/science.aat1378}

\bibitem[{{Jorstad} {et~al.}(2013){Jorstad}, {Marscher}, {Smith}, {Larionov},
  {Agudo}, {Gurwell}, {Wehrle}, {L{\"a}hteenm{\"a}ki}, {Nikolashvili},
  {Schmidt}, {Arkharov}, {Blinov}, {Blumenthal}, {Casadio}, {Chigladze},
  {Efimova}, {Eggen}, {G{\'o}mez}, {Grupe}, {Hagen-Thorn}, {Joshi},
  {Kimeridze}, {Konstantinova}, {Kopatskaya}, {Kurtanidze}, {Kurtanidze},
  {Larionova}, {Larionova}, {Sigua}, {MacDonald}, {Maune}, {McHardy}, {Miller},
  {Molina}, {Morozova}, {Scott}, {Taylor}, {Tornikoski}, {Troitsky}, {Thum},
  {Walker}, {Williamson}, {Sallum}, {Consiglio}, \&
  {Strelnitski}}]{jorstad2013}
{Jorstad}, S.~G., {Marscher}, A.~P., {Smith}, P.~S., {et~al.} 2013, \apj, 773,
  147, \dodoi{10.1088/0004-637X/773/2/147}

\bibitem[{{Keivani} {et~al.}(2018){Keivani}, {Murase}, {Petropoulou}, {Fox},
  {Cenko}, {Chaty}, {Coleiro}, {DeLaunay}, {Dimitrakoudis}, {Evans}, {Kennea},
  {Marshall}, {Mastichiadis}, {Osborne}, {Santander}, {Tohuvavohu}, \&
  {Turley}}]{Keivani18}
{Keivani}, A., {Murase}, K., {Petropoulou}, M., {et~al.} 2018, \apj, 864, 84,
  \dodoi{10.3847/1538-4357/aad59a}

\bibitem[{{Kirk} {et~al.}(2000){Kirk}, {Guthmann}, {Gallant}, \&
  {Achterberg}}]{Kirk00}
{Kirk}, J.~G., {Guthmann}, A.~W., {Gallant}, Y.~A., \& {Achterberg}, A. 2000,
  \apj, 542, 235, \dodoi{10.1086/309533}

\bibitem[{{Laing}(1980)}]{Laing80}
{Laing}, R.~A. 1980, \mnras, 193, 439, \dodoi{10.1093/mnras/193.3.439}

\bibitem[{{Li} {et~al.}(2018){Li}, {Guo}, {Li}, \& {Li}}]{Li18}
{Li}, X., {Guo}, F., {Li}, H., \& {Li}, S. 2018, \apj, 866, 4,
  \dodoi{10.3847/1538-4357/aae07b}

\bibitem[{{MAGIC Collaboration} {et~al.}(2018){MAGIC Collaboration}, {Ahnen},
  {Ansoldi}, {Antonelli}, {Arcaro}, {Baack}, {Babi{\'c}}, {Banerjee},
  {Bangale}, {Barres de Almeida}, {Barrio}, {Becerra Gonz{\'a}lez}, {Bednarek},
  {Bernardini}, {Ch Berse}, {Berti}, {Bhattacharyya}, {Biland}, {Blanch},
  {Bonnoli}, {Carosi}, {Carosi}, {Ceribella}, {Chatterjee}, {Colak}, {Colin},
  {Colombo}, {Contreras}, {Cortina}, {Covino}, {Cumani}, {da Vela}, {Dazzi},
  {de Angelis}, {de Lotto}, {Delfino}, {Delgado}, {di Pierro},
  {Dom{\'{\i}}nguez}, {Dominis Prester}, {Dorner}, {Doro}, {Einecke},
  {Elsaesser}, {Fallah Ramazani}, {Fern{\'a}ndez-Barral}, {Fidalgo}, {Fonseca},
  {Font}, {Fruck}, {Galindo}, {Gallozzi}, {Garc{\'{\i}}a L{\'o}pez},
  {Garczarczyk}, {Gaug}, {Giammaria}, {Godinovi{\'c}}, {Gora}, {Guberman},
  {Hadasch}, {Hahn}, {Hassan}, {Hayashida}, {Herrera}, {Hose}, {Hrupec},
  {Ishio}, {Konno}, {Kubo}, {Kushida}, {Kuve{\v z}di{\'c}}, {Lelas},
  {Lindfors}, {Lombardi}, {Longo}, {L{\'o}pez}, {Maggio}, {Majumdar},
  {Makariev}, {Maneva}, {Manganaro}, {Mannheim}, {Maraschi}, {Mariotti},
  {Mart{\'{\i}}nez}, {Masuda}, {Mazin}, {Mielke}, {Minev}, {Miranda},
  {Mirzoyan}, {Moralejo}, {Moreno}, {Moretti}, {Nagayoshi}, {Neustroev},
  {Niedzwiecki}, {Nievas Rosillo}, {Nigro}, {Nilsson}, {Ninci}, {Nishijima},
  {Noda}, {Nogu{\'e}s}, {Paiano}, {Palacio}, {Paneque}, {Paoletti}, {Paredes},
  {Pedaletti}, {Peresano}, {Persic}, {Prada Moroni}, {Prandini}, {Puljak},
  {Garcia}, {Reichardt}, {Rhode}, {Rib{\'o}}, {Rico}, {Righi}, {Rugliancich},
  {Saito}, {Satalecka}, {Schweizer}, {Sitarek}, {{\v S}nidari{\'c}},
  {Sobczynska}, {Stamerra}, {Strzys}, {Suri{\'c}}, {Takahashi}, {Takalo},
  {Tavecchio}, {Temnikov}, {Terzi{\'c}}, {Teshima}, {Torres-Alb{\`a}},
  {Treves}, {Tsujimoto}, {Vanzo}, {Vazquez Acosta}, {Vovk}, {Ward}, {Will},
  {Zari{\'c}}, {Fermi-Lat Collaboration}, {Bastieri}, {Gasparrini}, {Lott},
  {Rani}, {Thompson}, {MWL Collaborators}, {Agudo}, {Angelakis}, {Borman},
  {Casadio}, {Grishina}, {Gurwell}, {Hovatta}, {Itoh}, {J{\"a}rvel{\"a}},
  {Jermak}, {Jorstad}, {Kopatskaya}, {Kraus}, {Krichbaum}, {Kuin},
  {L{\"a}hteenm{\"a}ki}, {Larionov}, {Larionova}, {Lien}, {Madejski},
  {Marscher}, {Myserlis}, {Max-Moerbeck}, {Molina}, {Morozova}, {Nalewajko},
  {Pearson}, {Ramakrishnan}, {Readhead}, {Reeves}, {Savchenko}, {Steele},
  {Tornikoski}, {Troitskaya}, {Troitsky}, {Vasilyev}, \& {Zensus}}]{magic2018}
{MAGIC Collaboration}, {Ahnen}, M.~L., {Ansoldi}, S., {et~al.} 2018, \aap, 619,
  A45, \dodoi{10.1051/0004-6361/201832677}

\bibitem[{{Maraschi} {et~al.}(1992){Maraschi}, {Ghisellini}, \&
  {Celotti}}]{maraschi92}
{Maraschi}, L., {Ghisellini}, G., \& {Celotti}, A. 1992, \apjl, 397, L5,
  \dodoi{10.1086/186531}

\bibitem[{{Marscher}(2014)}]{marscher2014}
{Marscher}, A.~P. 2014, \apj, 780, 87, \dodoi{10.1088/0004-637X/780/1/87}

\bibitem[{{Marscher} \& {Gear}(1985)}]{marscher85}
{Marscher}, A.~P., \& {Gear}, W.~K. 1985, \apj, 298, 114,
  \dodoi{10.1086/163592}

\bibitem[{{Marscher} {et~al.}(2008){Marscher}, {Jorstad}, {D'Arcangelo},
  {Smith}, {Williams}, {Larionov}, {Oh}, {Olmstead}, {Aller}, {Aller},
  {McHardy}, {L{\"a}hteenm{\"a}ki}, {Tornikoski}, {Valtaoja}, {Hagen-Thorn},
  {Kopatskaya}, {Gear}, {Tosti}, {Kurtanidze}, {Nikolashvili}, {Sigua},
  {Miller}, \& {Ryle}}]{marscher2008}
{Marscher}, A.~P., {Jorstad}, S.~G., {D'Arcangelo}, F.~D., {et~al.} 2008, \nat,
  452, 966, \dodoi{10.1038/nature06895}

\bibitem[{{Marscher} {et~al.}(2010){Marscher}, {Jorstad}, {Larionov}, {Aller},
  {Aller}, {L{\"a}hteenm{\"a}ki}, {Agudo}, {Smith}, {Gurwell}, {Hagen-Thorn},
  {Konstantinova}, {Larionova}, {Larionova}, {Melnichuk}, {Blinov},
  {Kopatskaya}, {Troitsky}, {Tornikoski}, {Hovatta}, {Schmidt}, {D'Arcangelo},
  {Bhattarai}, {Taylor}, {Olmstead}, {Manne-Nicholas}, {Roca-Sogorb},
  {G{\'o}mez}, {McHardy}, {Kurtanidze}, {Nikolashvili}, {Kimeridze}, \&
  {Sigua}}]{marscher2010}
{Marscher}, A.~P., {Jorstad}, S.~G., {Larionov}, V.~M., {et~al.} 2010, \apjl,
  710, L126, \dodoi{10.1088/2041-8205/710/2/L126}

\bibitem[{{McEnery}(2019)}]{amego2019}
{McEnery}, J. 2019, in American Astronomical Society Meeting Abstracts, Vol.
  233, American Astronomical Society Meeting Abstracts \#233, \#158.22

\bibitem[{{M{\"u}cke} \& {Protheroe}(2001)}]{mucke01}
{M{\"u}cke}, A., \& {Protheroe}, R.~J. 2001, Astroparticle Physics, 15, 121,
  \dodoi{10.1016/S0927-6505(00)00141-9}

\bibitem[{{Nalewajko}(2017)}]{nalewajko2017}
{Nalewajko}, K. 2017, Galaxies, 5, 64, \dodoi{10.3390/galaxies5040064}

\bibitem[{{Nishikawa} {et~al.}(2005){Nishikawa}, {Hardee}, {Richardson},
  {Preece}, {Sol}, \& {Fishman}}]{Nishikawa05}
{Nishikawa}, K.-I., {Hardee}, P., {Richardson}, G., {et~al.} 2005, \apj, 622,
  927, \dodoi{10.1086/428394}

\bibitem[{{Ogle} {et~al.}(2005){Ogle}, {Davis}, {Antonucci}, {Colbert},
  {Malkan}, {Page}, {Sasseen}, \& {Tornikoski}}]{3c120_paper}
{Ogle}, P.~M., {Davis}, S.~W., {Antonucci}, R.~R.~J., {et~al.} 2005, \apj, 618,
  139, \dodoi{10.1086/425894}

\bibitem[{{Orienti} {et~al.}(2013){Orienti}, {Koyama}, {D'Ammando},
  {Giroletti}, {Kino}, {Nagai}, {Venturi}, {Dallacasa}, {Giovannini},
  {Angelakis}, {Fuhrmann}, {Hovatta}, {Max-Moerbeck}, {Schinzel}, {Akiyama},
  {Hada}, {Honma}, {Niinuma}, {Gasparrini}, {Krichbaum}, {Nestoras},
  {Readhead}, {Richards}, {Riquelme}, {Sievers}, {Ungerechts}, \&
  {Zensus}}]{orienti2013}
{Orienti}, M., {Koyama}, S., {D'Ammando}, F., {et~al.} 2013, \mnras, 428, 2418,
  \dodoi{10.1093/mnras/sts201}

\bibitem[{{Paliya} {et~al.}(2018){Paliya}, {Zhang}, {B{\"o}ttcher}, {Ajello},
  {Dom{\'{\i}}nguez}, {Joshi}, {Hartmann}, \& {Stalin}}]{Paliya18}
{Paliya}, V.~S., {Zhang}, H., {B{\"o}ttcher}, M., {et~al.} 2018, \apj, 863, 98,
  \dodoi{10.3847/1538-4357/aad1f0}

\bibitem[{{Petropoulou} {et~al.}(2015){Petropoulou}, {Dimitrakoudis},
  {Padovani}, {Mastichiadis}, \& {Resconi}}]{Petropoulou15}
{Petropoulou}, M., {Dimitrakoudis}, S., {Padovani}, P., {Mastichiadis}, A., \&
  {Resconi}, E. 2015, \mnras, 448, 2412, \dodoi{10.1093/mnras/stv179}

\bibitem[{{Raiteri} {et~al.}(2013){Raiteri}, {Villata}, {D'Ammando},
  {Larionov}, {Gurwell}, {Mirzaqulov}, {Smith}, {Acosta-Pulido}, {Agudo},
  {Ar{\'e}valo}, {Bachev}, {Ben{\'{\i}}tez}, {Berdyugin}, {Blinov}, {Borman},
  {B{\"o}ttcher}, {Bozhilov}, {Carnerero}, {Carosati}, {Casadio}, {Chen},
  {Doroshenko}, {Efimov}, {Efimova}, {Ehgamberdiev}, {G{\'o}mez},
  {Gonz{\'a}lez-Morales}, {Hiriart}, {Ibryamov}, {Jadhav}, {Jorstad}, {Joshi},
  {Kadenius}, {Klimanov}, {Kohli}, {Konstantinova}, {Kopatskaya}, {Koptelova},
  {Kimeridze}, {Kurtanidze}, {Larionova}, {Larionova}, {Ligustri}, {Lindfors},
  {Marscher}, {McBreen}, {McHardy}, {Metodieva}, {Molina}, {Morozova},
  {Nazarov}, {Nikolashvili}, {Nilsson}, {Okhmat}, {Ovcharov}, {Panwar},
  {Pasanen}, {Peneva}, {Phipps}, {Pulatova}, {Reinthal}, {Ros}, {Sadun},
  {Schwartz}, {Semkov}, {Sergeev}, {Sigua}, {Sillanp{\"a}{\"a}}, {Smith},
  {Stoyanov}, {Strigachev}, {Takalo}, {Taylor}, {Thum}, {Troitsky}, {Valcheva},
  {Wehrle}, \& {Wiesemeyer}}]{raiteri2013}
{Raiteri}, C.~M., {Villata}, M., {D'Ammando}, F., {et~al.} 2013, \mnras, 436,
  1530, \dodoi{10.1093/mnras/stt1672}

\bibitem[{{Rani} {et~al.}(2017){Rani}, {Krichbaum}, {Lee}, {Sokolovsky},
  {Kang}, {Byun}, {Mosunova}, \& {Zensus}}]{rani2017}
{Rani}, B., {Krichbaum}, T.~P., {Lee}, S.-S., {et~al.} 2017, \mnras, 464, 418,
  \dodoi{10.1093/mnras/stw2342}

\bibitem[{{Rani} {et~al.}(2015){Rani}, {Krichbaum}, {Marscher}, {Hodgson},
  {Fuhrmann}, {Angelakis}, {Britzen}, \& {Zensus}}]{rani2015}
{Rani}, B., {Krichbaum}, T.~P., {Marscher}, A.~P., {et~al.} 2015, \aap, 578,
  A123, \dodoi{10.1051/0004-6361/201525608}

\bibitem[{{Rani} {et~al.}(2013){Rani}, {Krichbaum}, {Fuhrmann}, {B{\"o}ttcher},
  {Lott}, {Aller}, {Aller}, {Angelakis}, {Bach}, {Bastieri}, {Falcone},
  {Fukazawa}, {Gabanyi}, {Gupta}, {Gurwell}, {Itoh}, {Kawabata}, {Krips},
  {L{\"a}hteenm{\"a}ki}, {Liu}, {Marchili}, {Max-Moerbeck}, {Nestoras},
  {Nieppola}, {Quintana-Lacaci}, {Readhead}, {Richards}, {Sasada}, {Sievers},
  {Sokolovsky}, {Stroh}, {Tammi}, {Tornikoski}, {Uemura}, {Ungerechts},
  {Urano}, \& {Zensus}}]{rani2013}
{Rani}, B., {Krichbaum}, T.~P., {Fuhrmann}, L., {et~al.} 2013, \aap, 552, A11,
  \dodoi{10.1051/0004-6361/201321058}

\bibitem[{{Rani} {et~al.}(2018){Rani}, {Jorstad}, {Marscher}, {Agudo},
  {Sokolovsky}, {Larionov}, {Smith}, {Mosunova}, {Borman}, {Grishina},
  {Kopatskaya}, {Mokrushina}, {Morozova}, {Savchenko}, {Troitskaya},
  {Troitsky}, {Thum}, {Molina}, \& {Casadio}}]{rani2018}
{Rani}, B., {Jorstad}, S.~G., {Marscher}, A.~P., {et~al.} 2018, \apj, 858, 80,
  \dodoi{10.3847/1538-4357/aab785}

\bibitem[{{Reimer} {et~al.}(2018){Reimer}, {Boettcher}, \& {Buson}}]{reimer18}
{Reimer}, A., {Boettcher}, M., \& {Buson}, S. 2018, arXiv e-prints.
\newblock \doarXiv{1812.05654}

\bibitem[{{Romanova} \& {Lovelace}(1992)}]{romanova92}
{Romanova}, M.~M., \& {Lovelace}, R.~V.~E. 1992, \aap, 262, 26

\bibitem[{{Schinzel} {et~al.}(2012){Schinzel}, {Lobanov}, {Taylor}, {Jorstad},
  {Marscher}, \& {Zensus}}]{schinzel2012}
{Schinzel}, F.~K., {Lobanov}, A.~P., {Taylor}, G.~B., {et~al.} 2012, \aap, 537,
  A70, \dodoi{10.1051/0004-6361/201117705}

\bibitem[{{Sikora} {et~al.}(1994){Sikora}, {Begelman}, \& {Rees}}]{sikora94}
{Sikora}, M., {Begelman}, M.~C., \& {Rees}, M.~J. 1994, \apj, 421, 153,
  \dodoi{10.1086/173633}

\bibitem[{{Sironi} \& {Spitkovsky}(2009)}]{Sironi09}
{Sironi}, L., \& {Spitkovsky}, A. 2009, \apjl, 707, L92,
  \dodoi{10.1088/0004-637X/707/1/L92}

\bibitem[{{Sironi} \& {Spitkovsky}(2014)}]{Sironi14}
---. 2014, \apjl, 783, L21, \dodoi{10.1088/2041-8205/783/1/L21}

\bibitem[{{Sironi} {et~al.}(2013){Sironi}, {Spitkovsky}, \& {Arons}}]{sironi13}
{Sironi}, L., {Spitkovsky}, A., \& {Arons}, J. 2013, \apj, 771, 54,
  \dodoi{10.1088/0004-637X/771/1/54}

\bibitem[{{Spada} {et~al.}(2001){Spada}, {Ghisellini}, {Lazzati}, \&
  {Celotti}}]{Spada01}
{Spada}, M., {Ghisellini}, G., {Lazzati}, D., \& {Celotti}, A. 2001, \mnras,
  325, 1559, \dodoi{10.1046/j.1365-8711.2001.04557.x}

\bibitem[{{Spitkovsky}(2008)}]{Spitkovsky08}
{Spitkovsky}, A. 2008, \apjl, 673, L39, \dodoi{10.1086/527374}

\bibitem[{{Tavecchio} {et~al.}(2018){Tavecchio}, {Landoni}, {Sironi}, \&
  {Coppi}}]{tavecchio18}
{Tavecchio}, F., {Landoni}, M., {Sironi}, L., \& {Coppi}, P. 2018, \mnras, 480,
  2872, \dodoi{10.1093/mnras/sty1491}

\bibitem[{{Weisskopf}(2018)}]{ixpe}
{Weisskopf}, M. 2018, Galaxies, 6, 33, \dodoi{10.3390/galaxies6010033}

\bibitem[{{Werner} {et~al.}(2018){Werner}, {Uzdensky}, {Begelman}, {Cerutti},
  \& {Nalewajko}}]{Werner18}
{Werner}, G.~R., {Uzdensky}, D.~A., {Begelman}, M.~C., {Cerutti}, B., \&
  {Nalewajko}, K. 2018, \mnras, 473, 4840, \dodoi{10.1093/mnras/stx2530}

\bibitem[{{Zhang} \& {B{\"o}ttcher}(2013)}]{Zhang13}
{Zhang}, H., \& {B{\"o}ttcher}, M. 2013, \apj, 774, 18,
  \dodoi{10.1088/0004-637X/774/1/18}

\bibitem[{{Zhang} {et~al.}(2014){Zhang}, {Chen}, \& {B{\"o}ttcher}}]{Zhang14}
{Zhang}, H., {Chen}, X., \& {B{\"o}ttcher}, M. 2014, \apj, 789, 66,
  \dodoi{10.1088/0004-637X/789/1/66}

\bibitem[{{Zhang} {et~al.}(2016{\natexlab{a}}){Zhang}, {Deng}, {Li}, \&
  {B{\"o}ttcher}}]{Zhang16}
{Zhang}, H., {Deng}, W., {Li}, H., \& {B{\"o}ttcher}, M. 2016{\natexlab{a}},
  \apj, 817, 63, \dodoi{10.3847/0004-637X/817/1/63}

\bibitem[{{Zhang} {et~al.}(2016{\natexlab{b}}){Zhang}, {Diltz}, \&
  {B{\"o}ttcher}}]{Zhang16b}
{Zhang}, H., {Diltz}, C., \& {B{\"o}ttcher}, M. 2016{\natexlab{b}}, \apj, 829,
  69, \dodoi{10.3847/0004-637X/829/2/69}

\bibitem[{{Zhang} {et~al.}(2019){Zhang}, {Fang}, {Li}, {Giannios},
  {B{\"o}ttcher}, \& {Buson}}]{Zhang19}
{Zhang}, H., {Fang}, K., {Li}, H., {et~al.} 2019, arXiv e-prints,
  arXiv:1903.01956.
\newblock \doarXiv{1903.01956}

\bibitem[{{Zhang} {et~al.}(2017){Zhang}, {Li}, {Guo}, \& {Taylor}}]{Zhang17}
{Zhang}, H., {Li}, H., {Guo}, F., \& {Taylor}, G. 2017, \apj, 835, 125,
  \dodoi{10.3847/1538-4357/835/2/125}

\bibitem[{{Zhang} {et~al.}(2018){Zhang}, {Li}, {Guo}, \&
  {Giannios}}]{zhang2018}
{Zhang}, H., {Li}, X., {Guo}, F., \& {Giannios}, D. 2018, ApJL, 862, L25,
  \dodoi{10.3847/2041-8213/aad54f}

\end{thebibliography}

\end{document}